\documentclass[12pt]{aastex}
\usepackage{emulateapj5}

\newcommand{\ovi}{\ion{O}{6}}

\newcommand{\cii}{\ion{C}{2}}
\newcommand{\piv}{\ion{P}{4}}
\newcommand{\mh}{\hbox{H$_2$}}

\newcommand{\kms}{\hbox{${\rm km\:s^{-1}}$}}
\newcommand{\cps}{\hbox{${\rm ct\:s^{-1}}$}}
\newcommand{\ccms}{\hbox{${\rm ct\:cm^{-2}\:s^{-1}}$}}
\newcommand{\flux}{\hbox{${\rm erg\:cm^{-2}\:s^{-1}\:\AA^{-1}}$}}
\newcommand{\um}{$\rm \mu$m}

\shorttitle{Sahnow et al.}
\shortauthors{On-Orbit Performance of FUSE}

\begin{document}

    \title{On-Orbit Performance of the {\em Far Ultraviolet
        Spectroscopic Explorer} ({\em FUSE}) Satellite}

\author{D. J. ~Sahnow\altaffilmark{1}, 
H. W.~Moos\altaffilmark{1},
T. B.~Ake\altaffilmark{1},
J.~Andersen\altaffilmark{2},
B-G Andersson\altaffilmark{1},
M.~Andre\altaffilmark{1,3},
D.~Artis\altaffilmark{4},
A. F.~Berman\altaffilmark{1},
W. P.~Blair\altaffilmark{1},
K. R.~Brownsberger\altaffilmark{5},
H. M.~Calvani\altaffilmark{1},
P.~Chayer\altaffilmark{1,6},
S. J.~Conard\altaffilmark{1},
P. D.~Feldman\altaffilmark{1},
S. D.~Friedman\altaffilmark{1}, 
A. W.~Fullerton\altaffilmark{1,6},
G. A.~Gaines\altaffilmark{7},
W. C.~Gawne\altaffilmark{2},
J. C.~Green\altaffilmark{4},
M. A.~Gummin\altaffilmark{7},
T. B.~Jennings\altaffilmark{2},
J. B.~Joyce\altaffilmark{1},
M. E.~Kaiser\altaffilmark{1},
J. W.~Kruk\altaffilmark{1},
D. J.~Lindler\altaffilmark{8},
D.~Massa\altaffilmark{9},
E. M.~Murphy\altaffilmark{1},
W. R.~Oegerle\altaffilmark{1},
R. G.~Ohl\altaffilmark{1},
B. A.~Roberts\altaffilmark{1}, 
M. L.~Romelfanger\altaffilmark{1},
K. C.~Roth\altaffilmark{1},
R.~Sankrit\altaffilmark{1},
K. R.~Sembach\altaffilmark{1},
R. L.~Shelton\altaffilmark{1},
O. H. W.~Siegmund\altaffilmark{7},
C. J.~Silva\altaffilmark{2},
G.~Sonneborn\altaffilmark{10},
S. R.~Vaclavik\altaffilmark{2},
H. A.~Weaver\altaffilmark{1},
and E.~Wilkinson\altaffilmark{5}}

\altaffiltext{1}{Department of Physics \& Astronomy, The Johns
Hopkins University, Baltimore, MD  21218}
\altaffiltext{2}{Honeywell Technology Solutions, Inc., Columbia, MD  21045}
\altaffiltext{3}{Institut d'Astrophysique de Paris, INSU CNRS, 98 bis
Boulevard Arago, F-75014 Paris, France}
\altaffiltext{4}{The Johns
Hopkins University Applied Physics Laboratory, Laurel, MD  20723}
\altaffiltext{5}{Center for Astrophysics and Space Astronomy,
University of Colorado, Boulder, CO  80309}
\altaffiltext{6}{Primary affiliation:  Dept. of Physics \& Astronomy,
University of Victoria, P.O. Box 3055, Victoria, BC, V8W 3P6, Canada.}
\altaffiltext{7}{Space Sciences Laboratory, University of California,
Berkeley, Berkeley, CA 94720-7450}
\altaffiltext{8}{Advanced Computer Concepts, Inc., Code 681, NASA/GSFC, Greenbelt, MD 20771}
\altaffiltext{9}{Raytheon ITSS, Code 681, NASA/GSFC, Greenbelt, MD 20771}
\altaffiltext{10}{Laboratory for Astonomy and Solar Physics, Code 681,
NASA/GSFC, Greenbelt, MD 20771}

\begin{abstract}
     Launch of the {\em Far Ultraviolet Spectroscopic Explorer} ({\em FUSE}) has been followed by an
extensive period of calibration and characterization as part of the  preparation for normal satellite operations. Major
tasks carried out during this period include initial coalignment, focusing and characterization of the four instrument
channels, and a preliminary measurement of the resolution and throughput performance of the instrument. We
describe the results from this test program, and present preliminary estimates of the on-orbit performance of the {\em
FUSE} satellite based on a combination of this data and prelaunch laboratory measurements.
\end{abstract}

\keywords{instrumentation: spectrographs--ultraviolet: general---space vehicles}

\section{Introduction}

     The {\em Far Ultraviolet Spectroscopic Explorer} ({\em
FUSE}) is obtaining high  resolution, far ultraviolet spectra of
faint astronomical objects in the 905 -1187 \AA\ wavelength
range \citep{Moos00}. Details of the {\em FUSE} design and
the predicted performance based on preflight measurements
have been given previously \citep{Friedman99, Sahnow96}.
After an initial period of spacecraft checkout and on-orbit
guidance tests, the two far ultraviolet detectors were powered
on in 1999 August, and several months of checkout and
science verification activities began. This included evaluation
of the overall satellite performance, and preliminary
instrument characterization. Science operations began in 1999
October, but characterization activity will continue
throughout the three year mission with a decreasing
frequency. Results from these early investigations show that
the satellite is, with a few exceptions, performing quite well;
most measures show that the performance is at or near
preflight predictions, and adequate to meet the goals of the
mission.

     The FUSE design consists of four coaligned optical
channels, two of which have optics coated with SiC (SiC1 and
SiC2), and two coated with LiF over Al (LiF1 and LiF2).
Each channel is made up of a telescope primary mirror, a
focal plane assembly containing the spectrograph entrance
apertures, a holographically-ruled diffraction grating, and a
portion of a detector. The previous paper \citep{Moos00}
presents an overview of the {\em FUSE} mission, including
the scientific background and an overall description of the
instrument. This paper discusses the measured performance of
the {\em FUSE} satellite, as of 2000 February, with an
emphasis on the properties of the instrument which affect the
scientific data.

\section{Detector Performance}

\subsection{Design}

     The two {\em FUSE} detectors have been described
previously \citep{Siegmund97, Sahnow00}.  Each detector has
two microchannel plate (MCP) segments with a helical double
delay line (DDL) anode.  The front surface of each of the four
segments is coated with a KBr photocathode to obtain quantum
efficiencies of 14 - 30\% across the {\em FUSE} bandpass,
depending on the segment and the wavelength. The
detectors are windowless, but in order to preserve the
photocathode, the microchannel plate and anode assemblies
were enclosed inside a vacuum box for ground testing and
calibration. On-orbit, a mechanical door was opened once the
spectrograph cavity pressure reached an acceptably low level. 
High voltage operations of the two detectors began on 1999
August 13 and August 26.

     The size and location of pixels in DDL detectors are
not fixed, but are determined by timing and analog
measurements. Temperature changes can therefore cause
geometric shifts and distortions in the detector, which must be
accounted for in the data processing. In order to characterize
these changes, electronic stimulation pulses are introduced into
the preamplifiers of the electronics during the recording of
spectra and flat fields. These stimulation pulses are injected
into the position readout anode, where they are processed
through the entire readout electronics chain. In this way, image
effects due to temperature variations in the anode and
electronics can be accurately tracked. The nominal size of the
{\em FUSE} detector pixels is 6 \um\ in x, and 9 - 16 \um\ in y,
depending on segment.

\subsection{Flat Field and Signal to Noise}

     After photon statistics, the most important source of
noise in the data is pixel-to-pixel variations and other
features in the detector response. Long flat field exposures
obtained on the ground contain 40-100 photon events per pixel.
Since a spectral resolution element of 0.04~{\AA} represents a
sum over $\sim$6 x-pixels and $\sim$7--70 y-pixels
(depending on the astigmatic height of the spectrum), they can
support signal-to-noise ratios of 50--120. Variations in the
thermal conditions during these exposures may make them
unsuitable for on-orbit use, however. Lower signal-to-noise flat
fields have been obtained on-orbit with an onboard stimulation
lamp; they are used to monitor any gross changes in detector
response. 

     The ground-based flats showed a complex structure
from blocked pores and other MCP defects. The pixel-to-pixel
response also shows a moir\'{e}-fringe like  quasi-periodic
variation with a $\sim$50 \um\ scale caused by the slightly
different  geometric scales between the front, middle and rear
MCP plates in the stack \citep{tremsin99}. These features
also appear in the on-orbit flats, to the limits of the signal-to-
noise of these measurements, although a careful comparison
requires taking account of the differences in temperature
between the ground measurements and the on-orbit data.

     An alternative method for minimizing the effects of
fixed pattern noise in the data is to dither the spectrum by
moving it back and forth in the dispersion direction on the
detector. Using this technique, a given wavelength falls on
different detector pixels; addition of the coaligned spectra
then averages out the fixed pattern noise. Observations of
G191-B2B acquired employing this strategy recovered
signal-to noise ratios of $\sim$120 per spectral resolution
element in one-dimensional spectra, consistent with photon
statistics (see Figure 1).

\subsection{Background}

     The background, from a combination of radioactivity
in the MCPs \citep{Siegmund88}, cosmic rays, other high
energy particles, and scattered airglow, is approximately 0.8
\ccms\ on all four segments at night; no hot spots are seen.
Before launch, when the particle background was not present,
typical rates were $\sim$0.35 \ccms. Because there is no
shutter on the {\em FUSE} instrument, the detectors are
constantly collecting photons, primarily from airglow when
there is no target in the apertures. This contamination means it
is not possible to get an accurate background measurement
from the detectors in the region where the spectra fall. Thus,
unused regions of the detector are used to measure the
intrinsic detector background. Since there appears to be no
identifiable structure to the background, no spatial variation
has been assumed.

       Although scattered light from the target is small
\citep{Moos00}, stray light is present at several levels. A
vertical ``stripe'' of enhanced counts is present on one
detector segment; its intensity varies with the Lyman-$\beta$
airglow, so it is thought to be caused by light entering the
spectrograph from an unknown location. In addition, scattered
light causes an increase in the background by a factor of 2 - 3
for observations made during the day. These effects are at a
level such that they affect data for only the lowest flux targets.
Finally, weak scattered solar emission lines have been
identified in the SiC-channel spectra during the sunlit portion of the
orbit at some pointing orientations.

     For observations where minimizing the detector
background is more important than maximizing the number of
photons collected, detector pulse height thresholds can be
applied to the data to further decrease the background to
$\sim$0.5 \ccms\ or less. Data is taken onboard with very
limited pulse height thresholding, but upper and lower
thresholds can be applied during processing of photon list data
in order to exclude more of the background, if appropriate.

\subsection{Flux limitations}

     {\em FUSE} was designed to observe faint objects.
As a result of its high sensitivity, it is difficult to observe
objects as bright as those observed by {\em Copernicus}. In
fact, the bright limit of $1\times10^{-10}$ \flux established in
order to ensure long term health of the detectors precludes
observing any of the {\em Copernicus} targets by several orders of
magnitude. In addition, the data system was not designed for
high count rates ($\geq$32,000 \cps). Bright emission line
objects present a different problem, since they can easily
deplete charge from the MCPs in local regions, which could
lead to a permanent loss of sensitivity at those wavelengths. In
addition, the intrinsic dead time of the detector electronics
limits a single detector segment to a count rate of $\sim$33,000
\cps. Higher flux targets could be observed, but they would
have such low efficiency that the detector lifetime would be
unduly compromised for a limited benefit.

     Variations in detector gain with time are being
monitored as part of the normal characterization program. It is
expected that as the mission progresses, the gain will drop, and
the high voltage will be raised to compensate for the decreased
gain.

\subsection{Single Event Upsets}

     The only significant detector anomaly discovered on
orbit was the sensitivity of the electronics to single event upsets
(SEUs). When the detector electronics were powered on for the
first time several days after launch, the detector data processing
unit began reporting errors in the memory which stores the
code controlling the detector. Further investigation revealed
that the memory was  being corrupted by high energy particles
as the satellite passed through the South Atlantic Anomaly
(SAA). These SEUs, which now occur roughly once every
three days on each detector, have no effect on the science data,
but are a potential detector health and safety issue, since
corruption of the executing code could cause unpredictable
behavior. Although it is not possible to decrease their
frequency (and, in fact, the approaching solar maximum may
cause it to increase), their effect has been minimized by
developing a procedure by which the instrument flight
computer reloads the corrupted detector code whenever an
SEU is detected. Rarely, about once per month, a potentially
more vital part of memory is corrupted, causing the detector
to reboot and turn off the detector high voltage as a safety
measure. This currently requires ramping up the high voltage
via ground commands, which typically results in a one day
loss of data for that detector; this process is now being
automated to reduce the loss of observing time.

\section{Mechanical Performance}

     To compensate for expected changes of alignment in
orbit, mechanisms to adjust the mirrors and focal plane assemblies (FPAs) 
containing the spectrograph entrance apertures
were built into the FUSE design. A carbon fiber structure with
an extremely low coefficient of thermal expansion, and
athermal optical mounts were used to minimize mechanical
variations which might affect the ability to hold the optics
stable to several microns over a several meter distance. 

     On orbit measurements have shown that thermal
motions are larger than expected, causing small rotations of
both the gratings and mirrors. The rotation of the gratings
causes the spectra to move in two dimensions on the detectors
in a roughly sinusoidal fashion over the course of an orbit; if
uncorrected, this smears the spectra by up to 0.09 \AA,
depending on the channel. This motion is correlated with the
pointing of the satellite, and appears to be earth-driven.
Algorithms have been developed to minimize these shifts
during the pipeline processing of the data. At present, these
corrections decrease the amplitude of the motion to less than
0.015 \AA. Additional studies are underway with the goal of
improving this further.

     The mirror rotation, which makes it difficult to keep
the four channels coaligned, initially resulted in a
misalignment large enough for targets to drift out of the large
(30\arcsec $\times$30\arcsec) aperture in one or more
channels in many instances. Since that time, we have
empirically mapped its behavior and limited the frequency of
this occurrence to less than $\sim$10\%\ of observations. In
addition, a theory exists for the cause of this motion. An
understanding of the drift in more detail is now being
obtained in preparation for making regular
observations in the smaller apertures.

     A number of workarounds were developed for these
problems. These are primarily operational constraints
imposed in order to control the moderate thermal variations
seen by certain parts of the satellite. Characterization is not
complete, but it is believed that with the development of
thermal models and careful scheduling of observations, this can
be improved in the future, allowing reasonable efficiency in the
30\arcsec\ $\times$ 30\arcsec\ and 4\arcsec\ $\times$ 20\arcsec\
apertures.

\section{Optical Performance}

     Optimal optical performance of {\em FUSE} depends
on a narrow point spread function (PSF) from the telescopes
and thus good slit transmission, proper coalignment of the four
channels, and a focused optical system. The focus problem for
each channel can
be divided into two parts: location of the optimal position of
the FPA in order to obtain
the best spectrograph focus, and proper telescope mirror to
FPA distance in order to maximize slit transmission. The FPAs
contain three apertures for general observing: a 30\arcsec\
$\times$ 30\arcsec\ aperture (LWRS) through which most
observations have been made thus far, an intermediate 4\arcsec\
$\times$20 \arcsec\ (MDRS) slit, and a narrow 1.25\arcsec\
$\times$ 20\arcsec\ slit (HIRS) aperture.

 \subsection{Telescope Performance}

     The telescopes provide a complex, non-uniform point
spread function at the FPAs. Preflight measurements and
analysis of a spare mirror, combined with metrology on the
flight mirrors, showed that 88$\pm$5\% of the encircled energy
is within a 1.5\arcsec\ diameter circle at 1000 \AA\
\citep{Ohl00}. In flight, knife edge scans made using the FPAs
showed that the telescope PSFs are consistent with these
ground measurements, with FWHMs of $\leq$1.1 - 1.7 arcsec,
depending on the channel \citep{OhlSPIE00}. This narrow
PSF means that the spectral resolution of point source objects
is not limited by the FPA apertures, but by the mechanical
stability factors described above; the $\sim$0.3\arcsec\ satellite
pointing stability also has a negligible effect.

     The knife edge scans mentioned above were also used
to optimize the mirror to FPA distances, which have been
located to $\pm$50 \um.

\subsection{Coalignment}

     Coarse alignment of the FUSE channels is achieved
by adjusting the telescope mirrors; this is required infrequently.
In the MDRS and HIRS apertures, fine alignment is
accomplished using the FPAs, by slewing the satellite,
measuring the signal in all four channels as a function of
position, and adjusting the FPA positions to maximize
throughput. In the LWRS aperture, these peakups are not
necessary.

     The ability to maintain coalignment of the four
channels is dependent on knowledge of the thermal and
mechanical stability of the system, which affects the rotation
of the telescope mirrors. Because of the mechanical
instabilities, observations in all but the LWRS aperture
require multiple peakups per orbit in order to maintain a
reasonable throughput. In the LWRS aperture, where
coalignment is much less critical, a satisfactory alignment can
typically be maintained for weeks without interruption. This
requires only small adjustments of the FPAs and mirrors as
long as satellite pointing angle constraints are followed.

\subsection{Spectrograph Resolution and Instrument Focus}

     The aberrations from a standard spherical grating
ruled with regularly spaced grooves would result in a point
source being imaged to a line many millimeters tall on the
detector. The holographic corrections in the {\em FUSE}
optical design \citep{Grange92, Green94} significantly
reduce the vertical extent of these images; they range from
$\sim$200 to $\sim$900 \um\ for a point source, depending on wavelength.
One of the astigmatic correction points, where the height is
minimized, was chosen to be at \ovi\ 1032 \AA\ for the LiF
channels, in order to limit the background contribution at this
astrophysically important line. The two correction points in
the SiC channels are near the edges of the bandpass, in order
to limit the height over as much of the spectrum as possible
and thus minimize the detector background. This size is
significant compared to the vertical extent of the entrance
apertures (220 - 330 \um), so that little or no spatial
information is available from the instrument, and then only at
wavelengths near the correction points. 

     A consequence of the holographic correction is that
the spectra include significant curvature due to astigmatism,
and the line width varies as a function of wavelength. As part
of the processing that occurs in the science data pipeline, this
curvature must be removed in order to achieve the highest
resolving power. At wavelengths away from a correction
point, there is a tradeoff between the amount of light included
(i.e. the vertical extent of the spectrum used in the analysis) 
and the resulting resolving power. The standard pipeline
extraction includes close to 100\% of light, but smaller
extraction windows can be used to obtain narrower line
spread functions at many wavelengths, although at the price of
lower throughput.

     The velocity resolution, $c\Delta\lambda/\lambda$, of
the {\em FUSE} spectrographs has  been determined by
measuring the widths of absorption features in the spectra of 
astrophysical objects. Figure 2 shows a plot in the 1041 - 1044
\AA\ region of the line of sight to WD0439+466 which has
been used to measure resolution. Since {\em FUSE} is limited
to observing faint objects, the velocity structure along the line
of sight to these objects is often complex, which can
significantly broaden the lines. Thus, these measurements
represent an upper limit to the instrument resolution.
Measurements with the spectrograph  focus settings at the
launch values showed upper limits to the velocity resolutions of 
$\sim$20 \kms, and values of this order are appropriate for the
interpretation of data presented in the accompanying papers.
Starting in 1999 December, adjustments were made to  the
instrument to optimize the resolving power. Those changes
have improved the SiC channels to $\sim$17 \kms, and the LiF
channels to $\sim$13 \kms\ in the LWRS.  Changes in
resolution as a function of  wavelength, which are on the order of 10\%,
are caused by both the
optical design and by variations of the intrinsic detector PSF
with position (typically $\sim$25 \um\ in the dispersion
direction). Before launch, measurements made using a \mh\
source showed marginally better resolution
\citep{Wilkinson98, Cha99}. Additional adjustments are
planned in order to improve these values further.

     On orbit, the positions of only the primary mirrors and
the FPAs can be adjusted. In the LWRS aperture, where most
of the early observations have been made, the focus  of the
instrument is determined solely by the position of the mirrors,
and the resolution is limited by the telescope PSF, the satellite
pointing stability, and the mirror and grating stability.
Observations in the smaller apertures, where the positions of
the FPAs are important,  have begun as our understanding of
the thermally-induced mirror motions has progressed.

     The highest possible resolution will require use of the
narrowest apertures, but this will likely result in a significant
loss of throughput, particularly in the SiC channels.

\section{On orbit calibration}

\subsection{Sensitivity}

     Observations of hot DA white dwarfs have
been used by previous  missions to obtain a photometric
calibration in this spectral region  \citep{dav92,
kru95,kru97,hur98,kru99}.  The preliminary {\em FUSE}
sensitivity  calibration used for early data analysis was
derived by two means: comparison of an observation of the
hot DA white dwarf WD2211-495 with a synthetic spectrum,
and comparison of an observation of  CSPN K1-16 with a
spectrum from the Hopkins Ultraviolet Telescope. The
synthetic spectrum of WD2211-495 was computed by D.
Finley, using the model  code of D. Koester, with an effective
temperature of 64000 K, log g=7.4, and  normalized to
V=11.77 \citep{fin97}.  Better-characterized  DA white
dwarfs (G191-B2B, HZ 43, and GD 246) became observable
later in the mission, which allowed a more accurate
determination of the effective area. The current estimate of
the effective area, based primarily on these stars, is shown in
Figure 3 for each of the channels as a function of wavelength.
It is comparable to preflight estimates.

      As part of our regular calibration program, we will
return to these calibration targets roughly once per month in
order to monitor the expected decrease in effective area due
to contamination and exposure to atomic oxygen
\citep{Moos00}. Early indications are that this drop is less than expected. 

     An optical anomaly, due to the astigmatism
in the system and a wire mesh in front of the detector, is present in the
data and can reduce the throughput over particular wavelength
intervals. The effect is visible in all channels,
but is most apparent in LiF1 above 1150 \AA\ (see Figure 3),
where it can reduce the flux by up to 40\%. Typically
the impact is much less, however. The position and severity depends
on observational parameters, such as the placement of the star in the
slit. After further characterization, the flux calibration will be
modified to account for this feature. 

\subsection{Wavelength Calibration}

     The {\em FUSE} instrument uses astrophysical
sources for wavelength calibration. Therefore, the absolute
wavelength determination requires comparison of spectra with
{\em HST} and ground-based data. During spectrograph
integration and test, \mh\ spectra were used to map the
wavelength scale, which is highly nonlinear because of the
variation in size of the analog detector pixels with position
\citep{Cha99}. The dispersion is 6.2 - 6.7 m\AA\ per pixel, depending
on the segment, with local variations of a few percent over most of
the detector.
High order polynomials were developed to describe this
distortion to approximately a spectrograph resolution element,
which was expected to remain stable, aside from the expected
stretches and shifts caused by variations in detector
temperature. A combination of downward looking airglow
and astronomical sources was planned to tie the ground-based
solution to the in flight data. However, these corrections have
not yet been fully implemented, limiting the accuracy of the
preliminary wavelength solution. It is generally better than 0.1
\AA, with some larger excursions. Activities currently
underway will substantially improve this.

\section{Event Bursts}

     An unexplained feature observed in the data is the
intermittent increase in the count rate, from an as yet
undetermined source. The pulse height distributions of these
``event bursts'' are consistent with the distributions exhibited
by photons, so they are apparently due to light rather than
particles or some other source internal to the detector. Their
spatial distribution on the detector does not match that of the
spectra, however. The source of these events is unknown.

     These bursts have durations that range from a few to
several hundred seconds, and maximum intensities that are
typically 20,000 per second. Attempts to correlate their
occurrence with orbital location, ram vector, or other orbital
phenomena have been unsuccessful, although they occur
primarily in orbital morning, with many occurring near noon.
Early in the mission, while the satellite was pointed in the
continuous viewing zone in order to avoid observing the bright
earth, they occurred on nearly every orbit. The frequency has
dropped significantly since that time, but it is unclear if that is
an effect of changes in pointing geometry due to constraints
provided by the mirror and grating motions, or some other
effect, such as a lower pressure in the spectrograph cavity.

     Since the bursts are isolated in time -- rarely occurring
more than once per orbit -- they can easily be screened from
time tag data during ground processing without a significant
loss of observing efficiency; a pipeline module to automatically
remove these times from the data is currently under
development. It is not possible to remove them from spectral
image data, but since these observations  typically have much
higher count rates, the bursts provide much less contamination.

\section{Summary}

     The {\em Far Ultraviolet Spectroscopic Explorer} is
performing well on orbit; early characterization activity is now
complete, and routine scientific observations have begun. We
have presented the first results from the characterization
program, which will continue throughout the mission. Despite
several as yet unexplained anomalies, high quality data
addressing a wide variety of scientific problems is being
collected, as seen in the accompanying papers.

\acknowledgments
     We gratefully acknowledge the many people who
have contributed to the success of the {\em FUSE} mission
during its design, construction and characterization. This work
is based on data obtained by the NASA-CNES-CSA {\em
FUSE} mission operated by the Johns Hopkins University.
Financial support to U. S. participants has been provided by
NASA contract NAS5-32985.

\clearpage

\figcaption[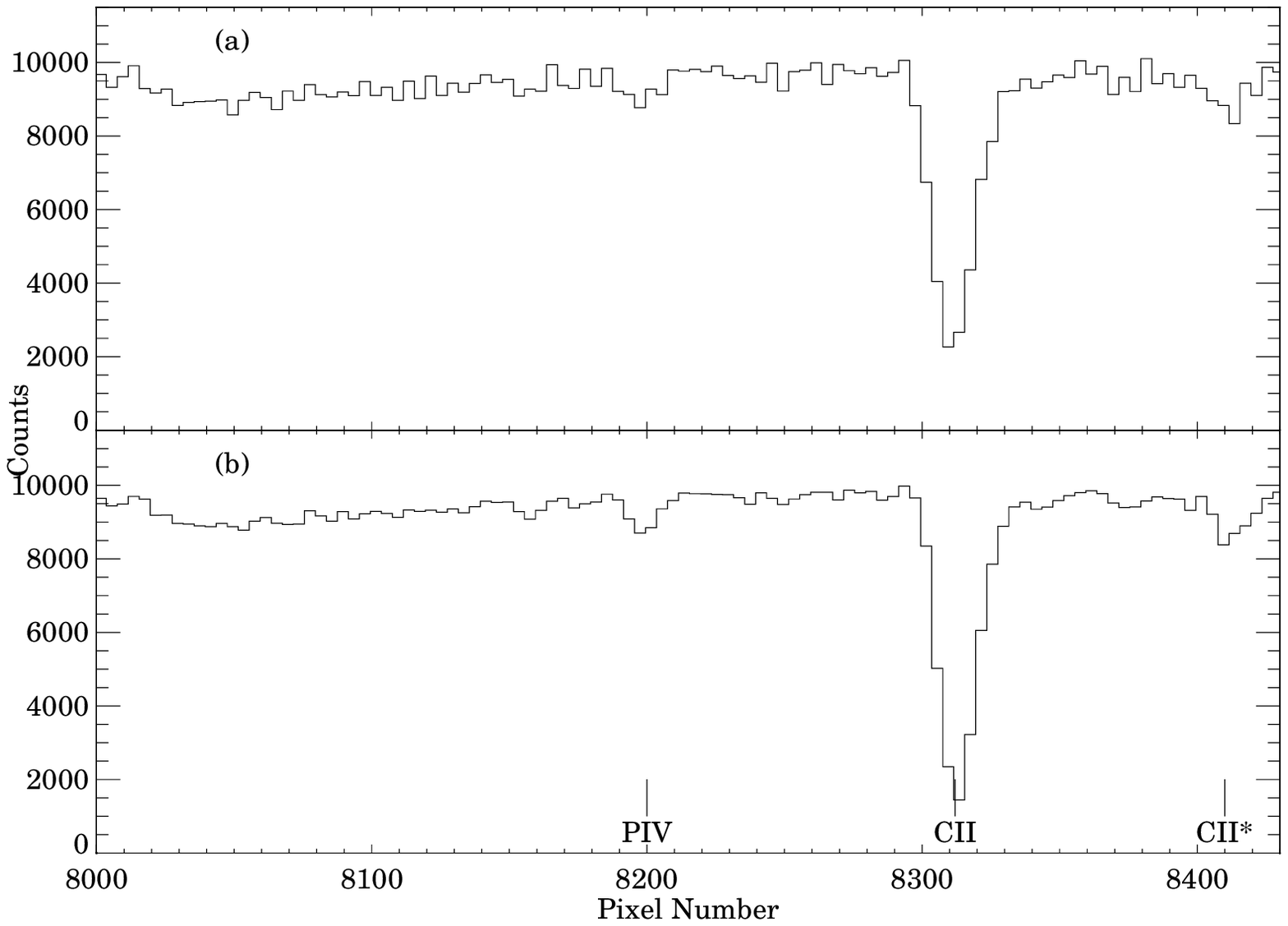]{A portion of the spectrum of G191-B2B,
binned by 4,
which includes spectral features from \cii\ $\rm\lambda$1036.34,
\cii* $\rm\lambda$1037.02 and \piv\ $\rm\lambda$1035.52.
Shown in (a) is an undithered spectrum, with a
signal-to-noise ratio of $\sim$63 per spectrograph resolution
element (6 pixels), while (b) displays a dithered spectrum, with a
signal-to-noise ratio of $\sim$120 per spectrograph resolution element.
The high frequency fixed pattern noise in the upper figure is
significantly reduced by dithering, so that the signal-to-noise
predicted by photon statistics is nearly recovered.\label{fig1}}

\figcaption[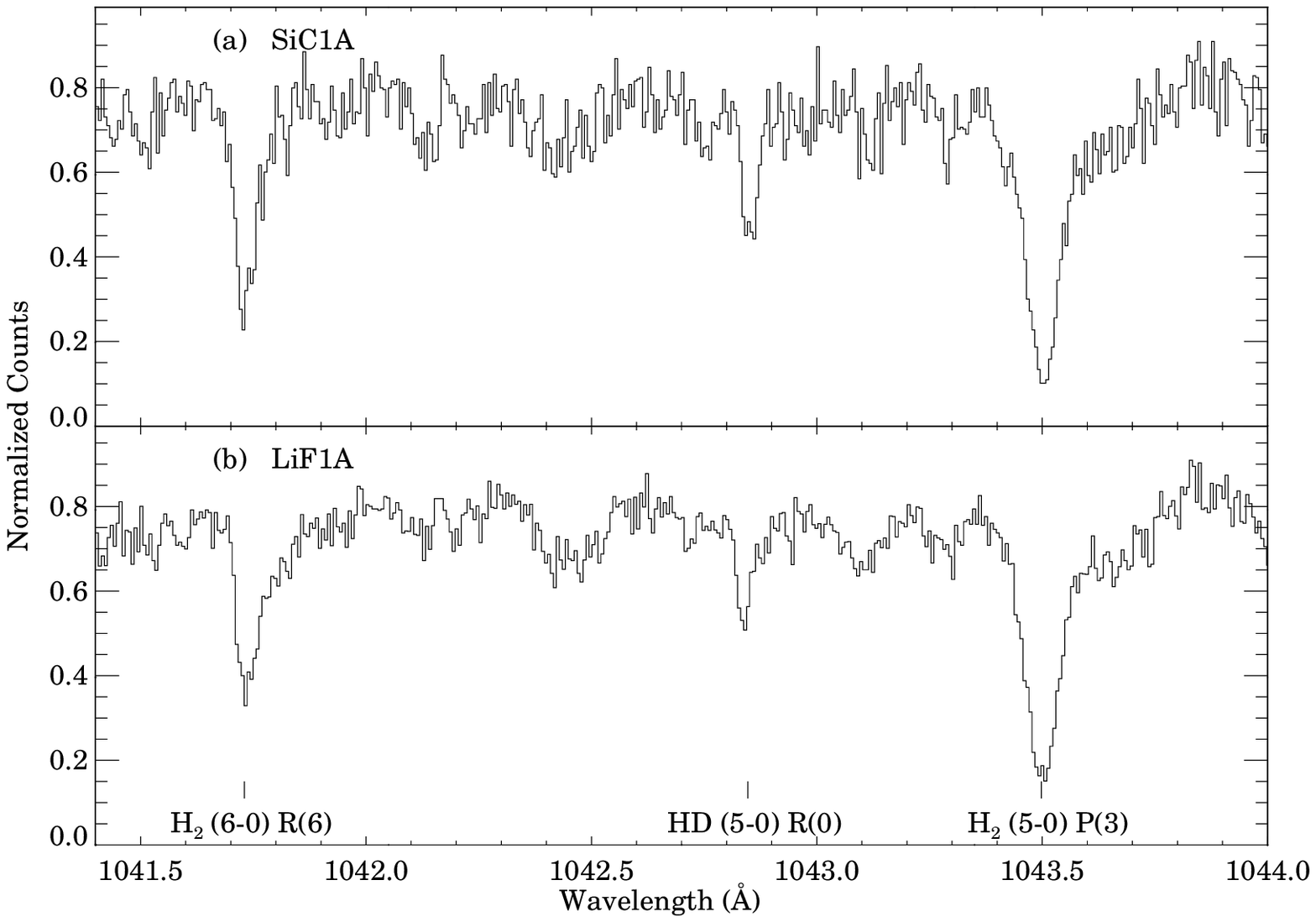]{A spectrum of the line of sight
towards WD0439+466 showing the (6-0)~R(6) and (5-
0)~P(3) transitions of \mh, and the (5-0)~R(0) line of HD in
the SiC1 and LiF1 channels. The HD line shows a resolving
power of $\sim$13 \kms. The unidentified features are probable
stellar absorption features.\label{fig2}}

\figcaption[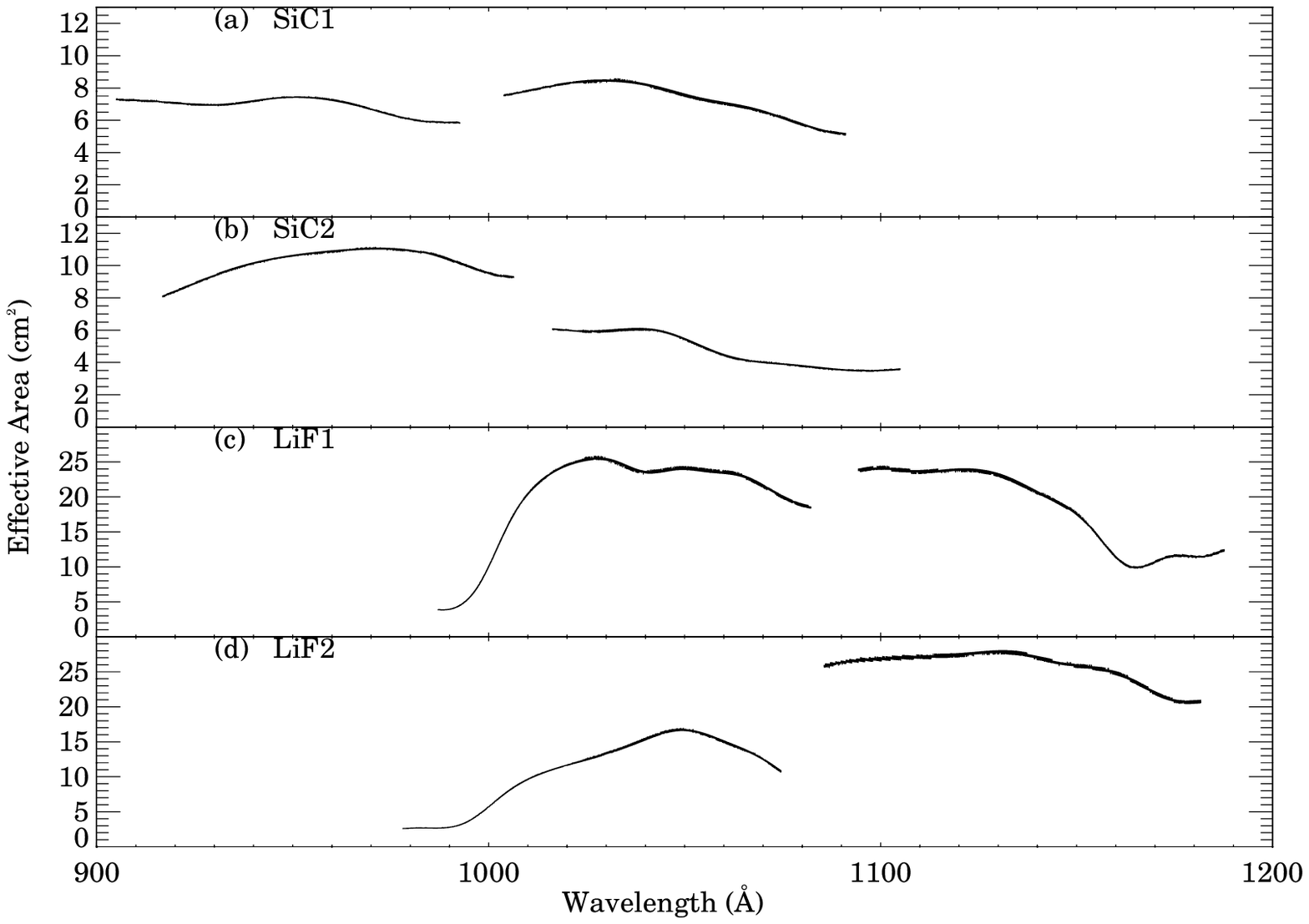]{Estimated effective area of the four
channels of the {\em FUSE} instrument, based on
measurements of white dwarfs and model spectra. Due to
variations in coating reflectivity, grating efficiency, and
detector sensitivity, there is a substantial variation between
channels. Errors are typically $\sim$10\%, except above
1150\AA. Note that the scales are different for the SiC and LiF
channels. \label{fig3}}

\begin{figure*}
\plotone{highsn.eps}
\end{figure*}

\begin{figure*}
\plotone{rescompare.eps}
\end{figure*}

\begin{figure*}
\plotone{aeff.eps}
\end{figure*}


\begin{thebibliography}{}
\bibitem[Cha, Sahnow \& Moos (1999)]{Cha99} Cha, A. N.,
Sahnow, D. J. \& Moos, H. W.  1999, \procspie,  3765, 495

\bibitem[Davidsen et al.(1992)]{dav92} Davidsen, A. F., et
al.  1992, \apjl,  392, L264

\bibitem[Finley et al.(1997)]{fin97} Finley, D.S., Koester, D.,
\& Basri, G.  1997, \apj, 488, 375

\bibitem[Friedman et al.(1999)]{Friedman99} Friedman, S.
D., et al. 1999, \procspie, 3765, 460 

\bibitem[Grange (1992)]{Grange92} Grange, R. 1992, \ao,
31, 3744 

\bibitem[Green et al.(1994)]{Green94} Green, J.C.,
Wilkinson, E. \& Friedman, S. D. 1994, \procspie, 2283, 12 

\bibitem[Hurwitz et al.(1998)]{hur98} Hurwitz, M., et al.
1998, \apj, 500, L1 

\bibitem[Kruk et al.(1995)]{kru95} Kruk, J. W., Durrance, S.
T., Kriss, G.  A., Davidsen, A. F., Blair, W. P., Espey, B. R.,
\& Finley, D. S. 1995,   \apj, 454, L1

\bibitem[Kruk et al.(1997)]{kru97} Kruk, J. W., Kimble, R.
A., Buss, R. H.,  Jr., Davidsen, A. F., Durrance, S. T., Finley,
D. S., Holberg, J. B., \&  Kriss, G. A. 1997, \apj, 482, 546

\bibitem[Kruk et al.(1999)]{kru99} Kruk, J. W., Brown, T.
M.., Davidsen, A.  F., Espey, B. R., Finley, D. S., \& Kriss, G.
A. 1999 \apjs, 122, 299  

\bibitem[Moos et al.(2000)]{Moos00} Moos, H. W. et al.
2000 \apjl, this volume

\bibitem[Ohl et al.(2000a)]{Ohl00} Ohl, R. G., Saha, T. T.,
Friedman, S. D., Barkhouser, R. H. \& Moos, H. W.  2000a,
\ao,  in press

\bibitem[Ohl et al.(2000b)]{OhlSPIE00} Ohl, R. G., et al.
2000b, \procspie, 4139

\bibitem[Sahnow et al.(1996)]{Sahnow96} Sahnow, D. J.,
Friedman, S. D., Oegerle, W. R., Moos, H. W., Green, J. C.
\& Siegmund, O. H. W. 1996, \procspie, 2807, 2 

\bibitem[Sahnow et al.(2000)]{Sahnow00} Sahnow, D. J., et al.
2000, \procspie, 4139

\bibitem[Siegmund, Vallerga \& Wargelin
(1988)]{Siegmund88} Siegmund, O. H. W., Vallerga, J. \&
Wargelin, B. 1988, IEEE~Trans.~Nucl.~Sci., 35, 524 

\bibitem[Siegmund et al.(1997)]{Siegmund97} Siegmund, O.
H. W., et al. 1997, \procspie, 3114, 283 

\bibitem[Tremsin et al.(1999)]{tremsin99} Tremsin, A. S.,
Siegmund, O. H. W., Gummin, M. A., Jelinsky, P. N. \& Stock,
J. M. 1999, \ao, 38, 2240 

\bibitem[Wilkinson et al.(1998)]{Wilkinson98} Wilkinson, E.,
Green, J. C., Osterman, S. N., Brownsberger, K. R. \& Sahnow,
D. J. 1998, \procspie, 3356, 18 

\end{thebibliography}
\end{document}